\documentclass[superscriptaddress,amsmath,amssymb,aps,floatfix,showpacs]{revtex4}
\usepackage[normalem]{ulem}
\usepackage{graphicx}
\usepackage{dcolumn}
\usepackage{bm}
\usepackage{hyperref}
\hypersetup{colorlinks=true,citecolor=blue,linkcolor=cyan,urlcolor=blue,filecolor= green, breaklinks=true}

\begin{document}

\title{Environment-induced quantum coherence spreading of a qubit}

\author{Mauro B. Pozzobom}

\address{Departamento de F\'isica, Centro de Ci\^encias Naturais e Exatas, Universidade Federal de Santa Maria, Avenida Roraima 1000, 97105-900, Santa Maria, RS, Brazil}

\author{Jonas Maziero}

\email{jonas.maziero@ufsm.br}

\address{Departamento de F\'isica, Centro de Ci\^encias Naturais e Exatas, Universidade Federal de Santa Maria, Avenida Roraima 1000, 97105-900, Santa Maria, RS, Brazil}

\address{Instituto de F\'isica, Facultad de Ingenier\'ia, Universidad de la Rep\'ublica, J. Herrera y Reissig 565, 11300, Montevideo, Uruguay}
\begin{abstract}
We make a thorough study of the spreading of quantum coherence (QC),
as quantified by the $l_{1}$-norm QC, when a qubit (a two-level quantum
system) is subjected to noise quantum channels commonly appearing
in quantum information science. We notice that QC is generally not
conserved and that even incoherent initial states can lead to transitory
system-environment QC. We show that for the amplitude damping channel
the evolved total QC can be written as the sum of local and non-local
parts, with the last one being equal to entanglement. On the other
hand, for the phase damping channel (PDC) entanglement does not account
for all non-local QC, with the gap between them depending on time
and also on the qubit's initial state. Besides these issues, the possibility
and conditions for time invariance of QC are regarded in the case
of bit, phase, and bit-phase flip channels. Here we reveal the qualitative
dynamical inequivalence between these channels and the PDC and show
that the creation of system-environment entanglement does not necessarily
imply in the destruction of the qubit's QC. We also investigate the
resources needed for non-local QC creation, showing that while the
PDC requires initial coherence of the qubit, for some other channels
non-zero population of the excited state (i.e., energy) is sufficient.
Related to that, considering the depolarizing channel we notice the
qubit's ability to act as a catalyst for the creation of joint QC
and entanglement, without need for nonzero initial QC or excited state
population.
\end{abstract}



\maketitle


\section{Introduction}

\label{sec_intro}

Although the main conceptual ideas related to quantum coherence (QC)
have been present in the literature for decades \cite{Feynman,Mandel-Wolf,Speekens_ijqi,Nori},
its formal quantification from a resource theory (RT) point of view
have been addressed only very recently \cite{Baumgratz,Aberg,Levi,Winter,Girolami_WY,Du_WY,Pati_ErQC,Adesso_QCoAss,Adesso_QCrob,Rana,Spekkens_RToC,Gour_RTC,Fan_Fid_TrD}.
Free states (FS) and free operations (FO) are two basic ingredients
in a RT \cite{Brandao_RTs}, with the resource states allowing us
to overcome the limitations imposed by the FO. Any function (leading
quantum states into non-negative real numbers) satisfying some basic
conditions and not increasing under the FO may be regarded as a monotone
measure for the resource under consideration. For instance, in the
RT of entanglement \cite{Plenio_E_rev}, the FS are the separable
ones while LOCC (local quantum operations and classical communication)
are the FO. Concurrence \cite{Wootters_EoF} and negativity \cite{Werner_En}
are two famous entanglement monotones; and we shall present their
definitions and use them in the next section.

In this article, we apply some developments achieved in the RT of
quantum coherence \cite{Baumgratz}. In this RT, the FS, the incoherent
states, are those density matrices which are diagonal in some reference
basis $|j\rangle$, i.e.,
\begin{equation}
\iota={\textstyle \sum_{j=1}^{d}}\iota_{j}|j\rangle\langle j|,\label{eq:iota}
\end{equation}
with $\iota_{j}$ being a probability distribution and $d=\dim\mathcal{H}$
is the dimension of the system's Hilbert space. We shall use $\mathcal{I}$
to denote the set of all density matrices of the type (\ref{eq:iota}).
The motivation to choose a particular reference basis usually originates
from the characteristic physical properties of the system under analysis.
For example, in transport phenomena in quantum biology, the eigenvectors
of the system's hamiltonian form a natural choice for the reference
basis \cite{Plenio_bookQB}. In this article, $|j\rangle$, with $j=0,1,\cdots,d-1$,
is always assumed to be the standard-computational basis in the regarded
Hilbert state.

Arguably, the most involved part of a RT is the identification of
its FO. While it is clear that the FO must not create resource states
from FS and should not increase the value of a resource quantifier
in the general case, which additional constraints one should add to
the FO is a subtle matter. The RT of coherence is a typical example
of this situation, where there exists several different levels of
restrictions one can impose on the FO \cite{Gour_RTC,Spekkens_RToC}.
In this work, we will use the $l_{1}$-norm quantum coherence, which
is defined and given by \cite{Baumgratz}:
\begin{equation}
C(\rho):=\min_{\iota\in\mathcal{I}}||\rho-\iota||_{l_{1}}={\textstyle \sum_{j\ne k}}|\langle j|\rho|k\rangle|,\label{eq:l1nc}
\end{equation}
where the $l_{1}$-norm of an operator $M$ is $||M||_{l_{1}}:=\sum_{j,k}|\langle j|M|k\rangle|$.
This coherence monotone satisfies most of the required properties.
For example, it is is zero if and only if $\rho$ is a incoherent
state and it does not increase under mixing or by (maximally) incoherent
operations. For more details, see Ref. \cite{Baumgratz}.

Recently, several articles have investigated the relation between
coherence (or other quantumnesses) and entanglement (or other quantum
correlations) and their inter-conversions via (incoherent) unitary
operations \cite{Adesso1,Plenio1,Vedral1,Leuchs2,Zubairy,Shahandeh,Girolami,Caves,Zhao,Fan}.
Here, we address a somewhat related question, but we regard the natural
dynamics generated due to the interaction of a two-level quantum system
(a qubit) with the environment. Besides its practical relevance regarding
the preservation of QC and its conversion to quantum correlations,
this kind of analysis is important e.g. to better understanding the
transition to classicality. This process is normally accompanied by
the loss of the system's QC, which is induced by the spontaneous creation
of correlations between it and its surroundings. It is worthwhile
mentioning that similar issues have been investigated previously,
but mainly with regard to decorrelating dynamics for composite systems
interacting with local or global environments, or in the thermodynamical,
non-Markovianity, and sensing contexts \cite{Maziero1,Maziero2,Walborn1,Fanchini1,Beims,Walborn2,Walborn3,Roszak,Fanchini2,Horodecki1,Leuchs1,Cao,Paternostro,Plenio2,Petruccione1}.
The goal of the present contribution is to provide a thorough description
of the dynamical flow of a quantum coherence monotone, the $l_{1}$-norm
QC, during the time evolution of a qubit interacting with environments
modeled by quantum channels important for quantum information science
\cite{Nielsen=000026Chuang,Wilde,Preskill}.

The reminder of this article is organized as follows. In Sec. \ref{sec_flow},
we start describing the initial conditions that shall be considered
in all subsequent calculations. Then we mention briefly the Kraus'
operator-sum representation and its corresponding unitary mapping.
Afterwards, we apply this formalism to study the $l_{1}$-norm quantum
coherence flow for a quantum bit evolving under the action of environments
modeled by the amplitude damping (Sec. \ref{sec_ad}), phase damping
(Sec. \ref{sec_pd}), bit flip (Sec. \ref{sec_bf}), phase flip (Sec.
\ref{sec_pf}), bit-phase flip (Sec. \ref{sec_bpf}), and depolarizing
(Sec. \ref{sec_d}) channels. Some final remarks about our findings
are included in Sec. \ref{sec_conclusions}.

\section{Dynamical flow of quantum coherence for quantum channels}

\label{sec_flow}

Throughout this article, we consider a qubit prepared initially in
a generic quantum state \cite{Nielsen=000026Chuang,Wilde}:
\begin{equation}
\rho^{S}=2^{-1}\left(\sigma_{0}+\vec{r}\cdot\vec{\sigma}\right).\label{eq:rho1qb}
\end{equation}
In the last equation, $\sigma_{0}$ is the identity matrix and the
components of the Bloch vector $\vec{r}=(r_{1},r_{2},r_{3})$ are
the system's polarizations: $r_{j}=\mathrm{Tr}(\rho^{S}\sigma_{j})$,
with $\sigma_{j}$ being the Pauli matrices. It is straightforward
seeing that the $l_{1}$-norm quantum coherence of a qubit-like system
represented by a Bloch vector $\vec{r}$ (as in Eq. (\ref{eq:rho1qb}))
is given by:
\begin{equation}
C(\rho^{S})=\sqrt{r_{1}^{2}+r_{2}^{2}}.
\end{equation}
The positivity of the density operator requires $\sum_{j=1}^{3}r_{j}^{2}\le1$,
which leads to $C(\rho^{S})\in[0,1]$. For a $d$-level system, a
qudit, we have $C(\rho^{S})\in[0,d-1]$ with the maximally coherent
state being $|\psi_{mc}\rangle=\sum_{j=0}^{d-1}|j\rangle/\sqrt{d}$.

After being prepared in the state (\ref{eq:rho1qb}), the qubit is
let to interact with the surroundings. We assume that initially the
two are uncorrelated, that the environment is in the vacuum state
$|E_{0}\rangle$, and that it is described by one of the quantum channels
regarded in the next sub-sections. Qubit plus environment form an
isolated system which evolves unitarily: $\rho_{t}=U_{t}(\rho^{S}\otimes|E_{0}\rangle\langle E_{0}|)U_{t}^{\dagger}$.
One can show that, if the matrix elements of the Kraus operators $K_{j}(t)$
are $\langle S_{l}|K_{j}(t)|S_{m}\rangle=(\langle S_{l}|\otimes\langle E_{j}|)U_{t}(|S_{m}\rangle\otimes|E_{0}\rangle)$,
the system's evolved state can be cast in the form \cite{Maziero_KR}:
\begin{equation}
\rho_{t}^{S}=\mathrm{Tr}_{E}(\rho_{t})={\textstyle \sum_{j}}K_{j}(t)\rho^{S}K_{j}^{\dagger}(t),\label{eq:KrausR}
\end{equation}
with $\mathrm{Tr}_{E}(\centerdot)$ being the partial trace function
\cite{Maziero_PTr}. 

If, as is most frequently the case, the Kraus' operators describing
the action of a noise channel are known from phenomenological or quantum
process tomography means \cite{Leung,Lidar}, we can model the system-environment
dynamics via the following isometric map \cite{Wilde}:
\begin{equation}
U_{t}|S_{l}\rangle\otimes|E_{0}\rangle={\textstyle \sum_{j}}(K_{j}(t)|S_{l}\rangle)\otimes|E_{j}\rangle,\label{eq:Umap}
\end{equation}
which also leads to the dynamics in Eq. (\ref{eq:KrausR}). It is
worthwhile mentioning that the set of Kraus' operators inducing a
certain evolution on the system state is not unique. In the context
we are interested in this article, $\{K_{l}(t)\}$ and $\{K_{j}'(t)=\sum_{l}V_{jl}K_{l}(t)\}$
lead to the same $\rho_{t}^{S}$ if $V$ is a unitary transformation
applied to the environment after its interaction with the system ceased
\cite{Maziero_KR}. As our main goal is to study the quantum coherence
spreading during and because of the system-environment interaction,
in the sequence we shall not regard this issue anymore. 

We observe that $|E_{k}\rangle$ are used to denote distinguishable
states of the environment on its $k$-excitations subspaces. Thus,
when we talk about the coherence of the environment and about the
system-environment coherence, we might be referring to subspaces,
instead of simply to some of the environment's bases states. In addition
to that, it is important stressing that although this formalism (Eq.
(\ref{eq:Umap})) does not utilize details about the environment's
``structure'', it is mathematically and physically correct and has
been applied in numerous previous works \cite{Maziero1,Maziero2,Walborn1,Walborn2,Walborn3,Walborn4,Davidovich,Eberly,Jing,Strunz}.
In the sequence we will apply it considering some relevant quantum
channels.

\subsection{Amplitude damping channel}

\label{sec_ad}

In this sub-section, we study a qubit embodied in a two-level atom-like
system\footnote{Another relevant context in which this channel appears is in quantum
communication with spin chains \cite{Bose_qcomm}.} with ground state $|0\rangle$ and excited state $|1\rangle$. In
the last equations and in the rest of this article, we use the notation:
\begin{equation}
|S_{j}\rangle\otimes|E_{k}\rangle=|jk\rangle.
\end{equation}
The environment is the electromagnetic field, which is initially in
the vacuum state (with no excitations). The amplitude damping channel
(ADC), represented by the Kraus' operators $K_{0}^{ad}(p)=|0\rangle\langle0|+\sqrt{q}|1\rangle\langle1|$
and $K_{1}^{ad}(p)=\sqrt{p}|0\rangle\langle1|$, is a phenomenological-approximate
representation of the spontaneous emission process, that will take
place because of the interaction of the atomic system with the vacuum
fluctuations \cite{Knight_book}. Above we defined $q:=1-p,$ with
$p\in[0,1]$ being the probability for the atom to emit a photon,
also dubbed \emph{parametrized time}. For instance, in liquid state
nuclear magnetic resonance (NMR), the explicit probability-time relation
is: $p=1-\exp(-t/T_{1})$, with $T_{1}$ being the so called longitudinal
relaxation time \cite{Maziero_NMR}. The unitary representation of
the ADC is \cite{Preskill}: $U_{ad}(p)|00\rangle=|00\rangle$ and
$U_{ad}(p)|10\rangle=\sqrt{1-p}|10\rangle+\sqrt{p}|01\rangle$. From
this unitary map, we obtain the system-environment evolved state 
\begin{eqnarray}
\rho_{p} & = & 2^{-1}[(1+r_{3})|00\rangle\langle00|+(1-r_{3})(p|01\rangle\langle01|+q|10\rangle\langle10|)\nonumber \\
 &  & \hspace{1em}\hspace{1em}+(1-r_{3})\sqrt{pq}|10\rangle\langle01|+(r_{1}-ir_{2})(\sqrt{p}|00\rangle\langle01|+\sqrt{q}|00\rangle\langle10|)+c.t.].
\end{eqnarray}
Above, and hereafter, we use $c.t.$ to denote the conjugate transpose
(adjoint) of the preceding term(s) in the parenthesis, bracket or
line (which one should be clear from the context). By using the partial
trace \cite{Maziero_PTr}, we can write the reduced states of the
system and environment as in Eq. (\ref{eq:rho1qb}), but with the
respective Bloch vector being:
\begin{align}
\vec{r}(p) & =\left(r_{1}\sqrt{q},r_{2}\sqrt{q},r_{3}q+p\right),\\
\vec{R}(p) & =\left(r_{1}\sqrt{p},r_{2}\sqrt{p},r_{3}p+q\right).
\end{align}

So the system-environment, system, and environment quantum coherences
at time $p$ are given, respectively, by:
\begin{align}
C(\rho_{p}) & =(\sqrt{q}+\sqrt{p})C(\rho^{S})+\sqrt{pq}(1-r_{3}),\label{eq:Cad}\\
C(\rho_{p}^{S}) & =\sqrt{q}C(\rho^{S}),\label{eq:CSad}\\
C(\rho_{p}^{E}) & =\sqrt{p}C(\rho^{S}).\label{eq:CEad}
\end{align}
Throughout this article, when computing the total coherence, we use
as reference basis the tensor product of local computational bases:
$\{|S_{j}\rangle\otimes|E_{k}\rangle\equiv|jk\rangle\}$, with $j=0,1$
and $k=0,\cdots,d_{E}-1$. We notice that the effective dimension
of the environment, $d_{E}$, is equal to the number of Kraus' operators
in a given representation of a quantum channel. 

The expressions in Eqs. (\ref{eq:Cad}), (\ref{eq:CSad}), and (\ref{eq:CEad})
show nicely the splitting of the total coherence into its \emph{local},
\begin{equation}
C_{l}(\rho_{p}):=C(\rho_{p}^{S})+C(\rho_{p}^{E}),
\end{equation}
and \emph{nonlocal},
\begin{equation}
C_{nl}(\rho_{p}):=C(\rho_{p})-C_{l}(\rho_{p}),
\end{equation}
parts\footnote{While completing this article we became aware of two related works:
Ref. \cite{Byrnes} investigated the distribution of multipartite
QC while Ref. \cite{Jeong} defined an entanglement monotone via the
NLQC of extensions of a bipartite state.}. We also see that QC is not conserved in general; note for instance
that $C_{l}(\rho_{p})=(\sqrt{q}+\sqrt{p})C(\rho^{S})\ge C(\rho^{S})$.
Besides, even if $C(\rho^{S})=0$, any initial state of the qubit
with non-null population of the excited state $|1\rangle$ (i.e.,
$r_{3}\ne1)$ shall lead to transitory system-environment non-local
QC, which will disappear only at the asymptotic time $p=1$.

For simplicity, in order to relate the non-local quantum coherence
with quantum correlations, let us begin by computing the \emph{concurrence}.
For two-qubit systems, this entanglement quantifier can be written
as \cite{Wootters_EoF}
\begin{equation}
E_{c}(\rho_{p})=\max\left(0,\sqrt{\lambda_{1}}-\sqrt{\lambda_{2}}-\sqrt{\lambda_{3}}-\sqrt{\lambda_{4}}\right),\label{eq:Ec}
\end{equation}
where $\lambda_{j}$ are the eigenvalues, in decreasing order, of
$\rho_{p}\tilde{\rho}_{p}$ with $\tilde{\rho}_{p}:=\sigma_{2}\otimes\sigma_{2}\rho_{p}^{*}\sigma_{2}\otimes\sigma_{2}$
and $\rho_{p}^{*}$ is the complex conjugate of $\rho_{p}$. One can
verify that $\rho_{p}\tilde{\rho}_{p}$ possess only one non-null
eigenvalue: $\lambda_{1}=pq(1-r_{3})^{2}$. Therefore, as $\lambda_{1}\ge0$,
we obtain $E_{c}(\rho_{p})=\sqrt{pq}(1-r_{3})=C_{nl}(\rho_{p}).$
Hence, we arrive at the following dynamical coherence-entanglement
relation:
\begin{equation}
C(\rho_{p})=C_{l}(\rho_{p})+E_{c}(\rho_{p}).
\end{equation}

We also notice that $C(\rho_{p=1}^{S})=E(\rho_{p=1})=0$ and $C(\rho_{p=1})=C(\rho_{p=1}^{E})=C(\rho^{S})$.
That is to say, for this channel all QC initially present in the qubit
is transferred to the environment in the asymptotic time ($p=1$).

\subsection{Phase damping channel}

\label{sec_pd}

The phase damping channel (PDC), which can be represented by the Kraus'
operators $K_{0}^{pd}(p)=\sqrt{q}\sigma_{0}$, $K_{1}^{pd}(p)=\sqrt{p}|0\rangle\langle0|$,
and $K_{2}^{pd}(p)=\sqrt{p}|1\rangle\langle1|$, is a phenomenological
description of the principal cause of coherence loss in several quantum
systems. The unitary map for this kind of qubit-environment interaction
is\footnote{We observe that the minimal isometric extension (Stinespring dilation)
\cite{Wilde} of the PDC uses only two Kraus' operators. However,
because of its nice phenomenological interpretation \cite{Preskill},
here we apply the isometric extension of the PDC with three Kraus'
operators.}: $U_{pd}(p)|00\rangle=\sqrt{q}|00\rangle+\sqrt{p}|01\rangle$ and
$U_{pd}(p)|10\rangle=\sqrt{q}|10\rangle+\sqrt{p}|12\rangle$. We see
from this map that the system base states do not change, i.e., it
does not exchange energy with the environment. However, each one of
these states leaves a unique ``fingerprint'' in the environment, causing
it to conditionally jump to a different configuration. Besides, here
the environment is modeled effectively as a three-level quantum system
(a qutrit). In the case of liquid state NMR, the environment can be
characterized by independent amplitude and phase damping channels.
The parametrized time for this last case is $p=1-\exp(-t/T_{2})$,
with the transverse relaxation time $T_{2}$ being usually much smaller
than the longitudinal relaxation time $T_{1}$.

For the same initial condition as stated above, the system-environment,
system, and environment evolved states are given by:
\begin{eqnarray}
\rho_{p} & = & 2^{-1}\{(1+r_{3})(q|00\rangle\langle00|+p|01\rangle\langle01|)+(1-r_{3})(q|10\rangle\langle10|+p|12\rangle\langle12|)\label{eq:rho_PD}\\
 &  & \hspace{1em}\hspace{1em}+\sqrt{pq}[(1+r_{3})|00\rangle\langle01|+(1-r_{3})|10\rangle\langle12|]+(r_{1}-ir_{2})(q|00\rangle\langle10|+p|01\rangle\langle12|)+c.t.\nonumber \\
 &  & \hspace{1em}\hspace{1em}+(r_{1}-ir_{2})\sqrt{pq}(|00\rangle\langle12|+|01\rangle\langle10|)+c.t.\},\nonumber \\
\rho_{p}^{S} & = & 2^{-1}\{(1+r_{3})|0\rangle\langle0|+(1-r_{3})|1\rangle\langle1|+[(r_{1}-ir_{2})q|0\rangle\langle1|+c.t.]\},\\
\rho_{p}^{E} & = & 2^{-1}\{2q|0\rangle\langle0|+p[(1+r_{3})|1\rangle\langle1|+(1-r_{3})|2\rangle\langle2|]+\sqrt{pq}[(1+r_{3})|0\rangle\langle1|+(1-r_{3})|0\rangle\langle2|+c.t.]\}.
\end{eqnarray}

Summing up the absolute value of the off-diagonal elements of these
three density matrices, we obtain their QCs:
\begin{align}
 & C(\rho_{p})=\left(1+2\sqrt{pq}\right)C(\rho^{S})+2\sqrt{pq},\label{eq:QC_SE_PDC}\\
 & C(\rho_{p}^{S})=qC(\rho^{S}),\label{eq:QC_S_PDC}\\
 & C(\rho_{p}^{E})=2\sqrt{pq}.\label{eq:QC_E_PDC}
\end{align}
We see that the PDC destroys the coherence of the system faster than
the ADC does, a fact well known in decoherence theory \cite{Zurek_td}.
Besides, for the PDC the QC of the environment is transient and does
not depend on the initial state of the qubit. We also notice that,
as $C(\rho_{p=1}^{S})=C(\rho_{p=1}^{E})=0$, any non-zero initial
QC present in the system shall be asymptotically ($p=1$) transformed
into non-local quantum coherence (NLQC).

\begin{figure}
\begin{centering}
\includegraphics[scale=0.47]{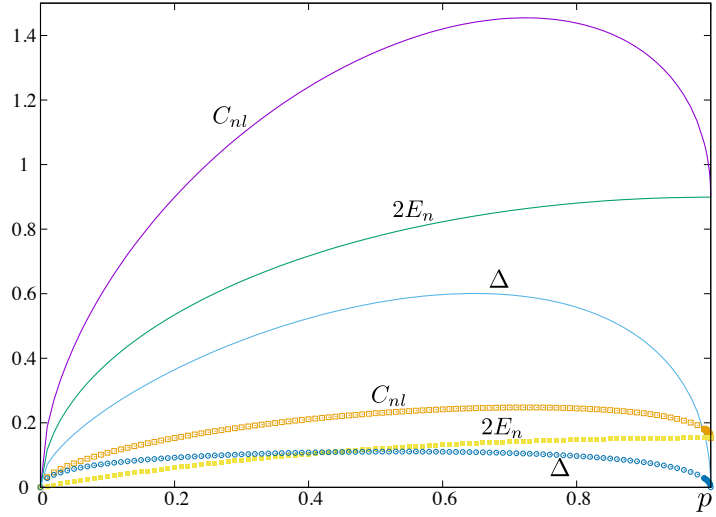}
\par\end{centering}

\caption{(color online) $l_{1}$-norm non-local quantum coherence, entanglement
negativity, and their gap as a function of the parametrized time $p$
for a qubit evolving under the phase damping channel. The lines and
points are obtained, respectively, for the qubit initial states $\vec{r}=(-0.41,0.80,-0.38)$
and $\vec{r}=(0.03,-0.15,-0.19)$.}

\label{fig:CnlEnt_PD}
\end{figure}

The next question we want to address is if, for the PDC, entanglement
is also, at least up to a constant factor, equivalent to this NLQC:
\begin{equation}
C_{nl}(\rho_{p})=\left(p+2\sqrt{pq}\right)C(\rho^{S}).
\end{equation}
As the Peres' criterion gives a necessary and sufficient condition
for separability of qubit-qutrit systems \cite{Peres_PPT,Horodecki_PPT},
for this channel we shall analyze the entanglement negativity, whose
definition is \cite{Werner_En}:
\begin{equation}
E_{n}(\rho_{p})=2^{-1}(||T_{E}(\rho_{p})||_{1}-1)=2^{-1}{\textstyle \sum_{j}}(|\lambda_{j}|-\lambda_{j}).\label{eq:En}
\end{equation}
In the last equation $T_{E}(\rho_{p})$ stands for the partial transpose
of $\rho_{p}$, $\lambda_{j}$ are the eigenvalues of $T_{E}(\rho_{p})$,
and $||X||_{1}=\mathrm{Tr}\sqrt{XX^{\dagger}}$ is the trace norm
\cite{Nielsen=000026Chuang}. In the general case, the characteristic
polynomial for $T_{E}(\rho_{p})$ is $\lambda^{2}(\lambda^{4}-\lambda^{3}+c_{2}\lambda^{2}+c_{1}\lambda+c_{0})=0$,
with $c_{2}=2^{-2}(1-r_{1}^{2}-r_{2}^{2}-r_{3}^{2})$, $c_{1}=2^{-2}p(2-p)(r_{1}^{2}+r_{2}^{2})$,
and $c_{0}=2^{-4}p^{2}(2-p)^{2}(r_{1}^{2}+r_{2}^{2})(r_{3}^{2}-1)$.
Although analytical expressions for the eigenvalues $\lambda_{j}$
can be obtained \cite{Shmakov}, they have a cumbersome form which
does not help in addressing the issue under analysis here. In contrast,
it is not difficult verifying that the equality $C_{nl}(\rho_{p=1})=2E_{n}(\rho_{p=1})$
holds asymptotically (and for $p=0$). Nevertheless, by generating
some random initial states $\rho^{S}$ \cite{Maziero_rrho_BJP,Maziero_FICT},
we checked that for $p\in(0,1)$ there exists a gap, $\Delta=C_{nl}-2E_{n}$,
between NLQC and entanglement which depends both on $p$ and on $\rho^{S}$.
Some examples are shown in Fig. \ref{fig:CnlEnt_PD}. Therefore, for
the PDC the coherence-entanglement relation is of the kind:
\begin{equation}
C(\rho_{p})=C_{l}(\rho_{p})+[2E_{n}(\rho_{p})+\Delta(p,\rho^{S})],
\end{equation}
with $\Delta(0,\rho^{S})=\Delta(1,\rho^{S})=0$. Moreover, we verified
that other quantum correlations \cite{Celeri_ijqi}, such as for example
the amended Hilbert-Schmidt quantum discord \cite{Akhtarshenas},
measurement-induced disturbance \cite{Luo_MID}, and measurement-induced
nonlocality \cite{Luo_MIN}, do worse than entanglement in accounting
for the PDC-induced NLQC. That is to say, in addition to show a different
qualitative behavior from the NLQC, these correlations cannot be made
to coincide with the NLQC even in the asymptotic time $p=1$. So,
we leave open the question of the possible description of this quantity
via a quantum correlation function.

\subsection{Bit flip channel}

\label{sec_bf}

The bit flip channel (BFC) is the most common error in classical information,
where a bit can be flipped, $0\leftrightarrows1$, due to random noise.
This type of state modification can also take place for a quantum
bit, where the computational base states can be left alone, $|0\rangle\rightarrow|0\rangle$
and $|1\rangle\rightarrow|1\rangle$, with probability $1-p$ or can
flipped, $|0\rangle\rightleftarrows|1\rangle$, with a probability
$p$. The Kraus operators for these transformations are \cite{Preskill}:
$K_{0}^{bf}(p)=\sqrt{q}\sigma_{0}$ and $K_{1}^{bf}(p)=\sqrt{p}\sigma_{1}$.
And the induced unitary map is: $U_{bf}(p)|00\rangle=\sqrt{q}|00\rangle+\sqrt{p}|11\rangle$
and $U_{bf}(p)|10\rangle=\sqrt{q}|10\rangle+\sqrt{p}|01\rangle.$
Using this map, we find the evolved states:
\begin{eqnarray}
\rho_{p} & = & 2^{-1}\{(1+r_{3})(q|00\rangle\langle00|+p|11\rangle\langle11|)+(1-r_{3})(q|10\rangle\langle10|+p|01\rangle\langle01|)\\
 &  & \hspace{1em}\hspace{1em}+\sqrt{pq}[(1+r_{3})|00\rangle\langle11|+(1-r_{3})|10\rangle\langle01|]+(r_{1}-ir_{2})(q|00\rangle\langle10|+p|11\rangle\langle01|)+c.t.\nonumber \\
 &  & \hspace{1em}\hspace{1em}+(r_{1}-ir_{2})\sqrt{pq}(|00\rangle\langle01|+|11\rangle\langle10|)+c.t.\},\nonumber \\
\rho_{p}^{S} & = & 2^{-1}\{[1+r_{3}(1-2p)]|0\rangle\langle0|+[1-r_{3}(1-2p)]|1\rangle\langle1|+[(r_{1}-ir_{2}(1-2p))|0\rangle\langle1|+c.t.]\},\\
\rho_{p}^{E} & = & q|0\rangle\langle0|+p|1\rangle\langle1|+r_{1}\sqrt{pq}(|0\rangle\langle1|+c.t.).
\end{eqnarray}

The formula for the total QC is equal to that obtained for the PDC
(Eq. (\ref{eq:QC_SE_PDC})). On the other hand, the local QCs are
given by:
\begin{eqnarray}
C(\rho_{p}^{S}) & = & \sqrt{r_{1}^{2}+r_{2}^{2}(1-2p)^{2}},\label{eq:QC_S_BFC}\\
C(\rho_{p}^{E}) & = & 2\sqrt{pq}|r_{1}|,\label{eq:QC_E_BFC}
\end{eqnarray}
which are not only quantitatively, but also qualitatively different
from those obtained for the ADC and PDC. Actually, for the BFC, the
distribution of QC is seen to be even more involved than for the previous
two channels. For instance, in contrast to the ADC and PDC, for the
BFC the QC can be made \emph{constant} (freezed) with time if the
qubit initial state has $r_{2}=0$. For a in-depth investigation of
frozen quantum coherence, see Ref. \cite{Adesso_frozen}. On the other
side, if $r_{1}=0$ the QC of the environment is always null and the
initial QC of the qubit is altogether transformed into NLQC (for $p=1$).
So, depending on $\rho^{S}$, the BFC can lead to qualitative behaviors
typical of the ADC or of the PDC, or to a ``mixture'' of their characteristics
traits.

\begin{figure}
\begin{centering}
\includegraphics[scale=0.47]{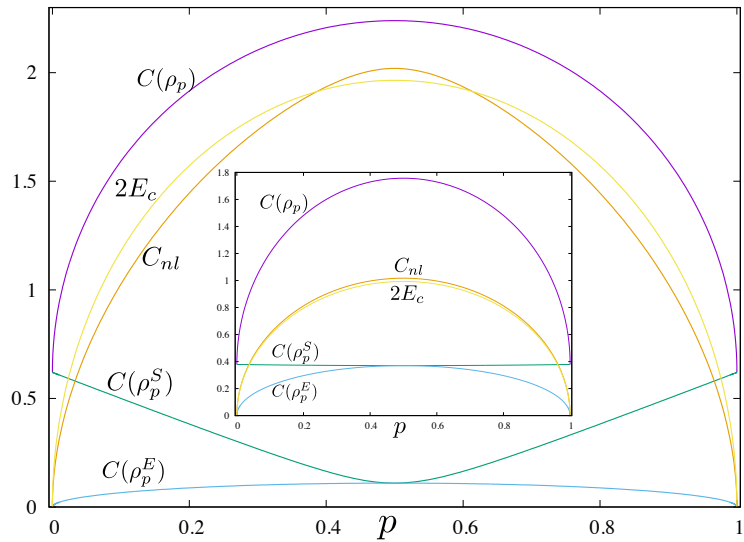}
\par\end{centering}

\caption{(color online) Coherences and entanglement for the time evolution
generated by the bit flip channel applied to a qubit prepared in the
state $\vec{r}=(-0.11,-0.61,0.77)$. The inset shows the same quantities
for $\vec{r}=(0.37,-0.08,-0.49)$. In this last case, even though
the quantum coherence of the system is almost constant, a considerable
amount of transitory non-local quantum coherence and entanglement
is generated. As seen in these figures and easily verified with Eqs.
(\ref{eq:QC_S_BFC}) and (\ref{eq:QC_E_BFC}), for any initial state,
at $p=1/2$ the coherences of the system and of the environment are
equal to $|r_{1}|$.}

\label{fig:bf}
\end{figure}

Now, let's analyze the NLQC and entanglement generated by the BFC.
Here, to compute the concurrence, we notice that the non-null eigenvalues
of $\rho_{p}\tilde{\rho}_{p}$ are: $\lambda_{\pm}=pq(\sqrt{x}\pm\sqrt{y})^{2}$,
where we used $x=r_{2}^{2}+r_{3}^{2}$ and $y=1-r_{1}^{2}$. So, the
concurrence of $\rho_{p}$ can be written as follows:
\begin{eqnarray}
E_{c}(\rho_{p}) & = & \sqrt{pq}\max\left(0,\sqrt{x}+\sqrt{y}-\left|\sqrt{x}-\sqrt{y}\right|\right)\nonumber \\
 & = & 2\sqrt{pq}\min(\sqrt{x},\sqrt{y}).\label{eq:E_BF}
\end{eqnarray}
By its turn, the created NLQC reads:
\begin{equation}
C_{nl}(\rho_{p})=(1+2\sqrt{pq})C(\rho^{S})-C(\rho_{p}^{S})+2\sqrt{pq}(1-|r_{1}|).
\end{equation}
Examples of the dynamics of coherences and entanglement are shown
in Fig. \ref{fig:bf}. As is indicated by the formulae, entanglement
is seem not to capture, in general, the quantitative behavior of the
NLQC generated by the BFC. Actually, the creation of system-environment
entanglement is shown here not to necessarily imply in the destruction
of the system's QC.

\subsection{Phase flip channel}

\label{sec_pf}

The phase flip channel (PFC) is a kind of error which only happens
in the quantum realm. The computational base states acquire random
phases differing by $\pi$. Or, disregarding global phases, one may
say that the state $|0\rangle$ is left alone (i.e., $|0\rangle\rightarrow|0\rangle$)
and the state $|1\rangle$ acquires a phase $\pi$ ($|1\rangle\rightarrow\mathrm{e}^{i\pi}|1\rangle$).
If the phase-flip error happens with a probability $p$, then we can
use the Kraus' operators $K_{0}^{pf}(p)=\sqrt{q}\sigma_{0}$ and $K_{1}^{pf}(p)=\sqrt{p}\sigma_{3}$,
which lead to the unitary map: $U_{pf}(p)|00\rangle=\sqrt{q}|00\rangle+\sqrt{p}|01\rangle$
and $U_{pf}(p)|10\rangle=\sqrt{q}|10\rangle-\sqrt{p}|11\rangle$.
The generated evolved density operators are:
\begin{eqnarray}
\rho_{p} & = & 2^{-1}\{(1+r_{3})(q|00\rangle\langle00|+p|01\rangle\langle01|)+(1-r_{3})(q|10\rangle\langle10|+p|11\rangle\langle11|)\\
 &  & \hspace{1em}\hspace{1em}+\sqrt{pq}[(1+r_{3})|00\rangle\langle01|-(1-r_{3})|10\rangle\langle11|]+(r_{1}-ir_{2})(q|00\rangle\langle10|-p|01\rangle\langle11|)+c.t.\nonumber \\
 &  & \hspace{1em}\hspace{1em}-(r_{1}-ir_{2})\sqrt{pq}(|00\rangle\langle11|-|01\rangle\langle10|)+c.t.\},\nonumber \\
\rho_{p}^{S} & = & 2^{-1}\{(1+r_{3})|0\rangle\langle0|+(1-r_{3})|1\rangle\langle1|+[(r_{1}-ir_{2})\left(1-2p\right)|0\rangle\langle1|+c.t.]\},\\
\rho_{p}^{E} & = & q|0\rangle\langle0|+p|1\rangle\langle1|+r_{3}\sqrt{pq}(|0\rangle\langle1|+c.t.).
\end{eqnarray}

From these density matrices, we compute the quantum coherences. For
the system we have
\begin{eqnarray}
C(\rho_{p}^{S}) & = & |1-2p|C(\rho^{S}).
\end{eqnarray}
The total coherence is once more equal to that obtained for the PDC
(Eq. (\ref{eq:QC_SE_PDC})). The QC of the environment is obtained
from the corresponding expression for the BFP, Eq. (\ref{eq:QC_E_BFC}),
by substituting $r_{1}$ with $r_{3}$.

\begin{figure}
\begin{centering}
\includegraphics[scale=0.47]{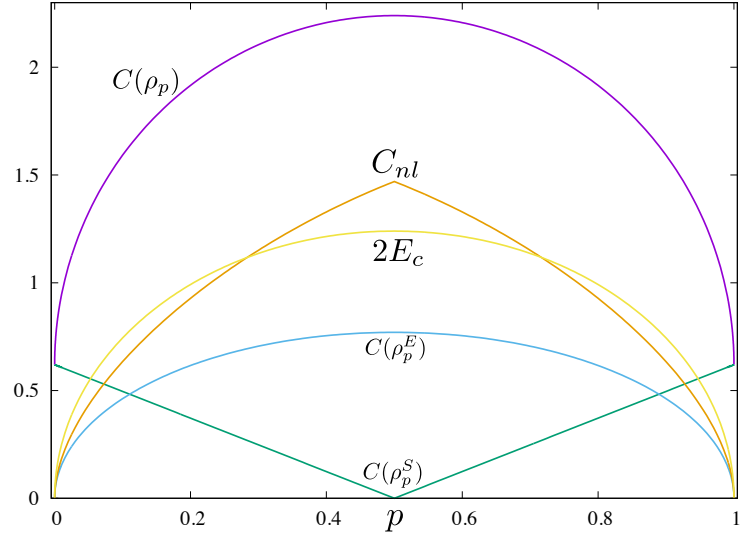}
\par\end{centering}

\caption{(color online) Coherences and entanglement for the time evolution
generated by the phase flip channel applied to a qubit prepared in
the state $\vec{r}=(-0.11,-0.61,0.77)$. In contrast to the previous
channels, for the PFC there is a discontinuity in the rate of change
of the system's and nonlocal quantum coherences at $p=1/2$. }

\label{fig:pf}
\end{figure}

We see thus that, in contrast to the BFC, for the PFC we cannot obtain
a constant value for the QC of the system. The qubit's QC monotonically
decreases up to $p=1/2$, which is the time (probability) for which
our uncertainty about what happened with the system is maximal. For
$p\in(1/2,1)$, the qubit's QC increases, going back to its initial
value when $p=1$. Besides, for the PFC the environment's QC does
not dependent on the initial QC of the system. In fact, the environment
gains transitory coherence whenever the qubit's ground and excited
states populations are not equal. We observe also that the concurrence
of $\rho_{p}$ for the PFC can be obtained from that in Eq. (\ref{eq:E_BF})
by the exchange of $r_{1}$ and $r_{3}$. The NLQC for the PFC is
\begin{equation}
C_{nl}(\rho_{p})=(1+2\sqrt{pq}-|1-2p|)C(\rho^{S})+2\sqrt{pq}(1-|r_{3}|).
\end{equation}
So, NLQC is created for all but the incoherent and excited state unpopulated
initial states. In Fig. \ref{fig:pf} is shown an example of the time
dependence of these coherences and entanglement.

\subsection{Bit-phase flip channel}

\label{sec_bpf}

The bit-phase flip channel (BPFC) is used to describe the situation
in which both the bit flip and the phase flip errors happen with probability
$p$, but simultaneously, i.e., $|0\rangle\rightarrow\mathrm{e}^{i\pi/2}|1\rangle$
and $|1\rangle\rightarrow\mathrm{e}^{i(-\pi/2)}|0\rangle$. To describe
this kind of noise interaction, we can use the Kraus operators: $K_{0}^{bpf}(p)=\sqrt{q}\sigma_{0}$
and $K_{1}^{bpf}(p)=\sqrt{p}\sigma_{2}$, which lead to the unitary
mapping: $U_{bpf}(p)|00\rangle=\sqrt{q}|00\rangle+i\sqrt{p}|11\rangle$
and $U_{bpf}(p)|10\rangle=\sqrt{q}|10\rangle-i\sqrt{p}|01\rangle$.
The density operators generated by this map are:
\begin{eqnarray}
\rho_{p} & = & 2^{-1}[(1+r_{3})(q|00\rangle\langle00|+p|11\rangle\langle11|)+(1-r_{3})(q|10\rangle\langle10|+p|01\rangle\langle01|)\\
 &  & \hspace{1em}\hspace{1em}+i\sqrt{pq}[(1-r_{3})|10\rangle\langle01|-(1+r_{3})|00\rangle\langle11|]+(r_{1}-ir_{2})(q|00\rangle\langle10|-p|11\rangle\langle01|)+c.t.\nonumber \\
 &  & \hspace{1em}\hspace{1em}+i(r_{1}-ir_{2})\sqrt{pq}(|00\rangle\langle01|+|11\rangle\langle10|)+c.t.],\nonumber \\
\rho_{p}^{S} & = & 2^{-1}\{[1+r_{3}(1-2p)]|0\rangle\langle0|+[1-r_{3}(1-2p)]|1\rangle\langle1|+[(r_{1}(1-2p)-ir_{2})|0\rangle\langle1|+c.t.]\},\\
\rho_{p}^{E} & = & q|0\rangle\langle0|+p|1\rangle\langle1|+r_{2}\sqrt{pq}(|0\rangle\langle1|+c.t.).
\end{eqnarray}

The expressions for the quantum coherences and entanglement associated
with these three density matrices are obtained from those in Eqs.
(\ref{eq:QC_SE_PDC}), (\ref{eq:QC_S_BFC}), (\ref{eq:QC_E_BFC}),
and (\ref{eq:E_BF}) by (when needed) exchanging $r_{1}$ and $r_{2}$.
Therefore, the analysis of the QC flow for the BPFC is similar to
that for the BFC. It is noticeable though that, in contrast to the
PFC, if the phase flip error happens accompanied by a bit flip error,
we regain the possibility for freezed QC, which is obtained for the
BPFC when $r_{1}=0$.

\subsection{Depolarizing channel}

\label{sec_d}

The depolarizing channel (DC) describes the situation in which the
interaction of the system with the surroundings mixes its state with
the maximally entropic one with a probability $p$, i.e., $\rho^{S}\mapsto(1-p)\rho^{S}+p\sigma_{0}/2.$
This kind of environment appears, for instance, in teleportation with
arbitrary mixed entangled resources \cite{Bose_Tel}. The DC can be
described using the following set of Kraus' operators: $K_{0}^{d}(p)=\sqrt{1-3p/4}\sigma_{0}$,
$K_{1}^{d}(p)=\sqrt{p/4}\sigma_{1}$, $K_{2}^{d}(p)=\sqrt{p/4}\sigma_{2}$,
and $K_{3}^{d}(p)=\sqrt{p/4}\sigma_{3}$, which lead to the unitary
map: $U_{d}(p)|00\rangle=\sqrt{1-3p/4}|00\rangle+\sqrt{p/4}(|11\rangle+i|12\rangle+|03\rangle)$
and $U_{d}(p)|10\rangle=\sqrt{1-3p/4}|10\rangle+\sqrt{p/4}(|01\rangle-i|02\rangle-|13\rangle)$.
So, this environment is modeled effectively as a four-level system
(a ququart). Now, if we define $u=p/4$ and $v=1-3u$, the evolved
states take the form:
\begin{eqnarray}
\rho_{p} & = & 2^{-1}\{(1+r_{3})[v|00\rangle\langle00|+u(|03\rangle\langle03|+|11\rangle\langle11|+|12\rangle\langle12|)]\\
 &  & \hspace{1em}\hspace{1em}+(1-r_{3})[v|10\rangle\langle10|+u(|01\rangle\langle01|+|02\rangle\langle02|+|13\rangle\langle13|)]\nonumber \\
 &  & \hspace{1em}\hspace{1em}+(1+r_{3})[\sqrt{uv}(|00\rangle\langle11|+|00\rangle\langle03|-i|00\rangle\langle12|)+u(|03\rangle\langle11|-i|03\rangle\langle12|-i|11\rangle\langle12|)]+c.t.\nonumber \\
 &  & \hspace{1em}\hspace{1em}+(1-r_{3})[\sqrt{uv}(|01\rangle\langle10|-i|02\rangle\langle10|-|10\rangle\langle13|)+u(i|01\rangle\langle02|-|01\rangle\langle13|+i|02\rangle\langle13|)]+c.t.\nonumber \\
 &  & \hspace{1em}\hspace{1em}+(r_{1}-ir_{2})[v|00\rangle\langle10|+\sqrt{uv}(|00\rangle\langle01|+i|00\rangle\langle02|-|00\rangle\langle13|+|11\rangle\langle10|+i|12\rangle\langle10|+|03\rangle\langle10|)]+c.t.\nonumber \\
 &  & \hspace{1em}\hspace{1em}+(r_{1}-ir_{2})u[|11\rangle(\langle01|+i\langle02|-\langle13|)+|12\rangle(i\langle01|-\langle02|-i\langle13|)+|03\rangle(\langle01|+i\langle02|-\langle13|)]+c.t.\},\nonumber \\
\rho_{p}^{S} & = & 2^{-1}\{(1+r_{3}q)|0\rangle\langle0|+(1-r_{3}q)|1\rangle\langle1|+[(r_{1}-ir_{2})q|0\rangle\langle1|+c.t.]\},\\
\rho_{p}^{E} & = & v|0\rangle\langle0|+u{\textstyle \sum_{j=1}^{3}}|j\rangle\langle j|+\sqrt{uv}({\textstyle \sum_{j=1}^{3}}r_{j}|0\rangle\langle j|+c.t.)+u[-i(r_{3}|1\rangle\langle2|-r_{2}|1\rangle\langle3|+r_{1}|2\rangle\langle3|)+c.t.].
\end{eqnarray}

\begin{figure}
\begin{centering}
\includegraphics[scale=0.47]{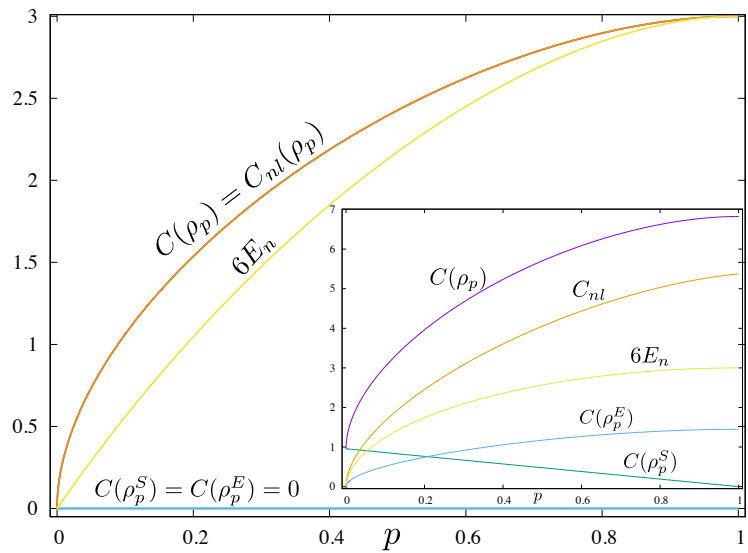}
\par\end{centering}

\caption{(color online) Dynamics of coherences and entanglement generated by
the depolarizing channel for the qubit initial state $\vec{r}=\vec{0}$.
The inset is for $\vec{r}=(-0.58,-0.76,0.11)$.}

\label{fig:dc}
\end{figure}

The total coherence can be written as follows:
\begin{equation}
C(\rho_{p})=(3\sqrt{u}+\sqrt{v})^{2}C(\rho^{S})+6(\sqrt{uv}+u).
\end{equation}
The expression for the QC of the system is equal to that for the PDC
(Eq. (\ref{eq:QC_S_PDC})). So the DC is seen to be as effective as
the PDC at erasing the QC of the qubit. The QC of the environment
at $p$ is given by:
\begin{equation}
C(\rho_{p}^{E})=2(\sqrt{uv}+u){\textstyle \sum_{j=1}^{3}|r_{j}|}.
\end{equation}
As $(\sqrt{uv}+u)_{p=1}=1/2$, we notice that, asymptotically, QC
is generated in the environment for all coherent and/or population
unbalanced initial states. For the DC the NLQC reads:
\begin{eqnarray}
C_{nl}(\rho_{p}) & = & [(3\sqrt{u}+\sqrt{v})^{2}-q]C(\rho^{S})+2(\sqrt{uv}+u)(3-{\textstyle \sum_{j=1}^{3}|r_{j}|}).
\end{eqnarray}
From these equations, we readily identify a somewhat puzzling phenomenon.
The DC, even for a maximally mixed or ground initial states of the
qubit, leads to the creation of NLQC (no energy or QC required). So,
for this channel, the system can be a catalyst whose state is not
modified but whose presence helps in creating joint QC and quantum
correlations. Examples of these dynamic behaviors are presented in
Fig. \ref{fig:dc}.

\section{Concluding Remarks}

\label{sec_conclusions}

In this article, we performed a detailed analysis of the dynamical
flow of the $l_{1}$-norm quantum coherence for a qubit interacting
with environments modeled by quantum channels relevant for quantum
information science. Our investigation provided several insights about
how these different kinds of system-environment interaction can affect
the quantum properties of the system, of the environment, and of their
correlations. We noticed that quantum coherence is not conserved in
general and that even incoherent initial states may lead to the creation
of transitory QC. For the amplitude damping channel, the non-local
QC created during the time evolution was shown to be completely captured
by the entanglement concurrence. This is not the case for the phase
damping channel. Besides, asymptotically the PDC transforms all initial
coherence of the qubit into non-local system-environment QC, while
the ADC only transfer it to the environment. The dynamic spreading
of QC due to the bit, phase, or bit-phase flip channels was shown
to be more diverse than for the ADC and PDC. Actually, by tuning the
initial state of the qubit we can observe dynamical behaviors typical
of these two channels or a mixture of them. We showed with this that,
contrary to what have been suggested previously (see e.g. Ref. \cite{Walborn4}),
the PDC is qualitatively distinct from the flip channels. Moreover,
for the flip channels, we have identified the possibility and conditions
for frozen QC and for a sudden modification in its rate of change
with time. Besides, the creation of system-environment entanglement
was shown here not necessarily to imply in the decaying of the system's
QC. We also investigated the qubit's initial resources needed for
creating non-local QC, showing that while the PDC requires QC the
ADC and phase flip channel need only nonzero population of the excited
state (i.e., energy). What is more, the potential of the qubit for
acting as a catalyst for the creation of joint QC and entanglement
by the depolarizing channel was reported.

Most of the literature on decoherence theory acknowledges that this
process takes place due to the creation of correlations between the
system and its surroundings. How much and what kind of correlation
is dynamically generated depends on the specificities of the system-environment
interaction Hamiltonian. Modeling these interactions in a general
setting is extremely difficult and surely out of reach. Quantum channels
are tools capable of providing such a description for a wide range
of physical situations. In this article, we used this tool to present
the most complete study of the environment-induced quantum coherence
flow up to now. Besides the several interesting findings described
in the main text and summarized in the last paragraph, our work may
contribute for two distinct directions of research. On one side, it
may be interesting to emulate the distinct dynamical behaviors induced
by distinct quantum channels in order to transform coherence into
quantum correlations in a prescribed and desired way. On the other
hand, understanding how the system loses its coherence and how this
coherence is transferred to the environment and/or transformed into
nonlocal properties (entanglement, discord, etc) is the essence of
what one might call understanding the decoherence process. Of course,
in quantum information science we would like to ultimately use this
knowledge to control, prevent or mitigate such kind of unwanted transformations.
We hope that our results and the solutions for the questions raised
here can shed some light for understanding and tackling this intricate
problem.
\begin{acknowledgments}
JM gratefully acknowledges the hospitality of the Physics Institute and Laser Spectroscopy Group at the Universidad de la Rep\'{u}blica, Uruguay.\\
\textbf{Funding}: This work was supported by the Brazilian funding agencies: Conselho Nacional de Desenvolvimento Cient\'ifico e Tecnol\'ogico (CNPq), processes 441875/2014-9 and 303496/2014-2, Instituto Nacional de Ci\^encia e Tecnologia de Informa\c{c}\~ao Qu\^antica (INCT-IQ), process 2008/57856-6, and Coordena\c{c}\~ao de Desenvolvimento de Pessoal de N\'{i}vel Superior (CAPES), process 6531/2014-08.\end{acknowledgments}

\bibliography{SEcohAOP}{}

\end{document}